\documentclass[]{aa}  
\usepackage{graphicx,epsfig}
\usepackage{txfonts}
\usepackage{natbib}
\usepackage{longtable}
\bibpunct{(}{)}{;}{a}{}{,} 
\begin{document}
  \title{A peculiar Of star in the Local Group galaxy IC\,1613\thanks{Based on observations obtained at 
  the ESO VLT for Programmes 078.D-0767 and 080.D-0423.}}

   \author{A. Herrero \inst{1,2}, M. Garcia \inst{1,2},  J. Puls \inst{3}, K. Uytterhoeven \inst{1,2}, F. Najarro \inst{4}, D.J. Lennon \inst{5} and J.G. Rivero-Gonz\'alez \inst{3}
	  }

   \offprints{A. Herrero (\email{ahd@iac.es})}

  \institute{Instituto de Astrof\'{i}sica de Canarias, C/ V\'{i}a L\'{a}ctea s/n, E-38200 La Laguna, Tenerife, Spain.
         \and
	 	Departamento de Astrof\'{i}sica, Universidad de La Laguna, Avda. Astrof\'{i}sico
		Francisco S\'anchez s/n, E-38071 La Laguna, Tenerife, Spain.
	\and
		Universit\"atssternwarte M\"unchen, Scheinerstr. 1, 81679 M\"unchen, Germany.	 \and
	 	Centro de Astrobiolog\'{i}a (CSIC-INTA), Ctra. de Torrej\'on a Ajalvir km-4, E-28850, Torrej\'on de Ardoz, Madrid, Spain.
	\and
		 ESA, Space Telescope Science Institute, 3700 San Martin Drive, Baltimore, MD 21218, USA.}

   \date{}

 
  \abstract
{Results from the theory of radiatively driven winds are nowadays incorporated in stellar evolutionary and population synthesis models, and are
used in our interpretation of the observations of the deep Universe. Yet, the theory has been confirmed only until Small Magellanic Cloud (SMC) 
metallicities. Observations
and analyses of O-stars at lower metallicities are difficult, but much needed to prove the theory.}
{We have observed GHV-62024, an O6.5 IIIf star in the low-metallicity galaxy IC\,1613 ($Z\approx0.15 Z_\odot$) to study its evolution and wind. 
According to a previous preliminary analysis that was subject to significant restrictions this star
could challenge the radiatively driven wind theory at low metallicities. 
Here we present a complete analysis of this star.}
{Our observations were obtained with VIMOS at VLT, at $R\approx2000$ and covered approximately between 4000 and
7000 \AA. The observations were analysed using the latest version of the model atmosphere code FASTWIND, which includes the possibility of
calculating the \ion{N}{iii} spectrum}
{We obtain the stellar parameters and conclude that the star follows the average wind momentum--luminosity relationship (WLR) expected for its 
metallicity, but with a high value for the exponent of the wind velocity law, $\beta$. Comparing this with values of
other stars in the literature, we suggest that this high value may be reached because GHV-62024 could be a fast rotator seen at a low inclination angle.
We also suggest that this could favour the appearance of the spectral "f"-characterictics. 
While the derived $\beta$ value does not change by adopting a lower wind terminal velocity, we show that a wrong
$V_\infty$ has a clear impact on the position of the star in the WLR diagram.
The \ion{N}{} and \ion{He}{} abundances are very high, consistent with strong CNO mixing that could have been
caused by the fast rotation, although we cannot discard a different origin with present data. 
Stellar evolutionary model predictions are consistent with the star being still a fast rotator. We find again the well-known mass-discrepancy
for this star.}
{We conclude that the star follows the WLR expected for its metallicity. The results are 
consistent with GHV-62024 being a fast rotator seen close to pole-on, strongly contaminated at the surface with CNO products and 
with a wind structure altered by the fast rotation but without modifying the global
WLR. We suggest that this could be a general property of fast rotators.}
  
	\authorrunning{A. Herrero et al.}
	\titlerunning{A peculiar Of star}
   \maketitle


\section{Introduction}

Three main parameters determine the structure and evolution of a star: its mass, angular momentum and metallicity. For massive stars the first two parameters
change rapidly during their lives because of stellar winds, whose strength is determined by the third parameter.
A reliable theory of stellar winds for massive stars is accordingly crucial for our predictions about
how these stars evolve and therefore for our predictions of their properties (ionizing flux, mechanical energy delivered, chemical composition, etc.)
at a given moment in their lives and final fate (type of supernova and remnant left behind). 

The theory of radiatively driven winds, used to explain the observed winds in massive stars and whose predictions are 
incorporated in stellar evolutionary and population synthesis models, is based on the work by \citet{cak75}, with improvements by 
\citet{ppk86} and \cite{friend86}. Recent reviews
can be found in \cite{kudritzki00} and \citet{puls08}. One of the most important predictions of the theory is that the modified wind momentum rate of the stellar
wind (MWM, the product of mass-loss rate, wind terminal velocity and square root of the stellar radius, which 
we denote as $D_\mathrm{mom}= \dot{M}v_\infty R^{1/2}$ following the use in the literature) should depend almost solely on the stellar luminosity 
and metallicity.  The VLT-FLAMES survey of massive stars 
(\citealt{evans05}, P.I. S. Smartt) was a project aimed precisely at testing this prediction, which was nicely confirmed by \citet{mokiem07b} by analysing 
O-type stars in the Milky Way, the LMC and the SMC. However, the second part of the project, the comparison of the predictions of stellar evolutionary
models including mass-loss and rotation with the stellar properties, revealed problems when comparing the nitrogen abundances (\citealt{hunter08}),
indicating that other physical processes may be at work. This started the VLT Tarantula FLAMES Survey (VFTS), a 
detailed investigation of the massive stellar population of 30 Doradus in the LMC currently under way \citep{evans11}. There a deeper 
study of the behaviour of the LMC O-stars winds will be possible, but without extending the available metallicity baseline.

Very many O-stars have been analysed in the Magellanic Clouds (see, among others, \citealt{bouret03}; \citealt{hillier03}; 
\citealt{mokiem06}; \citealt{mokiem07}; \citealt{massey04} or
\citealt{massey05}). However, going to lower metallicities implies a large distance jump. 
Yet, many star-forming galaxies in the Local Universe and presumably most stars in the early epochs of the Universe have metallicities below that of the
Small Magellanic Cloud. It is therefore natural to try to go one step further and analyse stars at even lower metallicities, studying their wind behaviour
and stellar evolution. Although going to galaxies farther away and with lower metallicity than the SMC will result in an increasing
difficulty to obtain good quality spectra that can be analysed, present day telescopes and instruments clearly allow for the next step. At metallicities of
about half that of the SMC and distances of one to a few megaparsecs it is still possible to perform reliable analyses. This has been done using
the visually brightest objects, the A- and B- supergiants, by \cite{urbaneja08} in the WLM (0.96 Mpc), \cite{evans07} in NGC\,3109 (1.3 Mpc), \cite{castro12}
in NGC\,55 (2.0 Mpc) or \cite{urbaneja05} in NGC\,300 (2.2 Mpc). \cite{bresolin01}
have even analysed individual A supergiants in NGC\,3621 at 6.7 Mpc. The hotter O-stars, however, are visually fainter and much more difficult
to observe at far distances. For this reason, there have been no spectroscopic analyses of 
individual O-stars located in low-Z galaxies using modern standard techniques  (i.e., analysing optical spectra by means of NLTE, spherical
models with mass-loss), not even in the Local Group, until very recently (\citealt{herrero11}; \citealt{tramper11}).
  
IC\,1613 is a dwarf irregular galaxy in the Local Group  with a distance modulus of $(m-M)_0 = 24.31\pm 0.06$
\citep{dolphin01}. It has a metallicity of $\log (O/H)+12 =  7.80 \pm 0.10$ as determined from its B-supergiants \citep{bresolin07}, 
whereas nebular studies vary between 7.60 and 7.90 \citep{lee03}. The metallicity of the B-supergiants 
indicates $Z_\mathrm{IC\,1613}= 0.13 \pm 0.07~ Z_\odot$ (using solar abundances from \citealt{asplund09}). This value is below the value derived by most
authors for the SMC O- and B-stars. For example, \citet{heap06} obtain $Z_\mathrm{SMC}= 0.2~ Z_\odot$ from an analysis of O-stars 
and \citet{dufton05} obtain $Z_\mathrm{SMC}= 0.25~ Z_\odot$ from an analysis of B-supergiants. 
The situation is similar for other authors,
e.g. \citet{bouret03}, who obtained $Z_\mathrm{SMC} = 0.2~ Z_\odot$. Therefore the stellar metallicity in IC\,1613 is below the SMC value
and the study of its stars represents a step forward in our analyses at low metallicities.

IC\,1613 shows a recent and intense burst of massive star formation, particularly in its 
NE part. We have recently published a new catalogue of 
OB associations in IC\,1613 (\citealt{garcia09}, identified as [GHV2009]) and their physical properties \citep{garcia10} as part of
our effort to carry out an in-depth study of the young population of IC\,1613. 

We also obtained spectra of some stars in this galaxy using VIMOS at VLT. In the field of IC\,1613 we found the star 62024 in the GHV2009 catalogue
(hereafter we identify the star as GHV-62024), an Of star. This star is ideally suited to study
the stellar evolution and winds in IC\,1613, because it displays strong \ion{N}{iii} lines and a P-Cygni profile in \ion{He}{ii} 4686.
A preliminary analysis of this star by \citet{herrero11} resulted in a too high log$D_{\rm mom}$ as expected from its metallicity and luminosity,
which would challenge the theory of radiatively driven winds at low metallicities. An even more serious challenge is the very recent
work by \cite{tramper11}. These authors analysed six O-stars in low-Z galaxies (four in IC\,1613 and one in WLM and NGC\,3109). The four stars
in IC\,1613 tend to show too high values of  $D_{\rm mom}$, although they are not inconsistent with the theory because of the large error bars. But the
two other stars have too high $D_{\rm mom}$ values, not consistent with theoretical predictions. However, although these two recent works 
seem to point towards a problem between observations and theory, we note that the analyses are subject to strong assumptions.
 
In the present work, we carry out a detailed analysis of GHV-62024, showing that when we drop the assumptions made by
\cite{herrero11}, we obtain a MWM value consistent with the theoretical expectations, although the
analysis opens alternative questions.
The observations and data reduction are presented in Sect.~\ref{obs}. The spectrum is described in Sect.~\ref{spec}. 
The spectral analysis follows in Sect.~\ref{anal} and a discussion of the stellar wind momentum rate is given in Sect.~\ref{discus}.
The evolutionary status and other properties are discussed in Sect.~\ref{evol}. Finally, the conclusions are presented in Sect.~\ref{conc}.


\section{Observations and data reduction}
\label{obs}

Details of the observations are given in \citet{herrero10}. We repeat here the main points.
Observations were performed with VIMOS at VLT in MOS (multi-object spectroscopy) mode.
The HR-blue and HR-orange gratings were used, resulting in a resolution of $R \sim 2050$
and $R \sim 2150$, respectively, and a spectral wavelength coverage roughly from 3870 to 7240 \AA~
(depending on the star position).
We had 19 observing blocks for the blue part of the spectrum, each with three individual exposures,
and ten observing blocks for the red part, each with two individual exposures. Thus, 
in total $3\times19$ blue spectra were observed from October 5 to November 5, 2007,
and $2\times10$ red spectra from November 5 to November 13, 2007. Integration times
for individual exposures were 650 sec for the blue spectra and 1005 to 1110 sec for the
red spectra. The three (two) consecutive exposures in the blue (red) were coadded. The
coadded spectra were extracted with standard IRAF\footnote{IRAF is
distributed by the National Optical Astronomy Observatory, which is operated by the
Association of Universities for
Research in Astronomy, Inc., under cooperative agreement with  the National Science
Foundation.} procedures after wavelength
calibration and cosmic ray removal. Next, a barycentric correction and a correction for
the systemic velocity of IC\,1613 ($-234  \pm1~ {\mathrm {km~ s^{-1} }}$, \citealt{lu93}) were applied. Finally, after checking that no significant radial velocity variations 
are present, all 19 (10) resulting spectra were
coadded to a single blue (red) spectrum. The resulting spectra, with an S/N of 85 both in the blue and red wavelength regions, were then rectified.

\begin{figure}
     \includegraphics[angle=0,width=9.0cm]{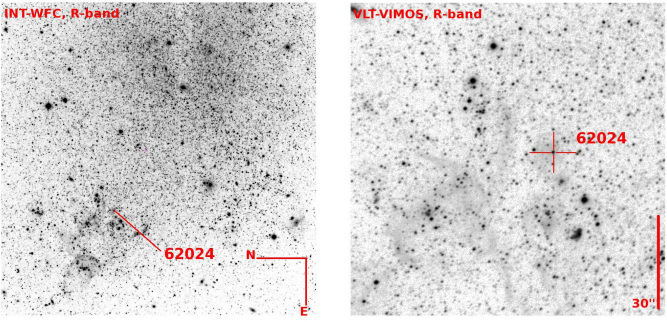}
     \caption{Finding chart for GHV-62024. The left part is an image obtained with the WFC at the INT in the R band. The region imaged (5x5 arcmin) 
     contains the IC\,1613 intense star-forming region, with many ionized bubbles. The right-hand side is a VLT-VIMOS image in the same band, at a larger scale
     (indicated by the scale bar at the right edge). The orientation is the same in both images.}
	\label{finding}
  \end{figure}

Figure~\ref{finding} shows the finding chart of GHV-62024. The star is located at $\alpha$(2000)= 01h 05m 00.72s, $\delta$(2000)= 2$^\circ$ 8$^{\prime}$ 49$^{\prime\prime}$.1 (from the astrometry by \citealt{garcia09}). It clearly lies in a zone of intense recent star formation, with strong ionized
bubbles visible in Fig.~\ref{finding}. Table~\ref{stardata} gives the optical photometric data of the star from the catalogue of \citet{garcia09}.

\begin{table}[htdp]
\caption{Photometric data for GHV-62024 from \citet{garcia09}.}
\centering
\begin{tabular}{ccccccc}
\hline \hline
 U & B & V & R & I \\
 \hline
 18.69  & 19.44  & 19.60 &  19.76 & 19.68  \\
\hline
\end{tabular}
\label{stardata}
\end{table}

\section{Description of the spectrum}
\label{spec}

\begin{figure*}
     \includegraphics[angle=90,height=8cm,width=15cm]{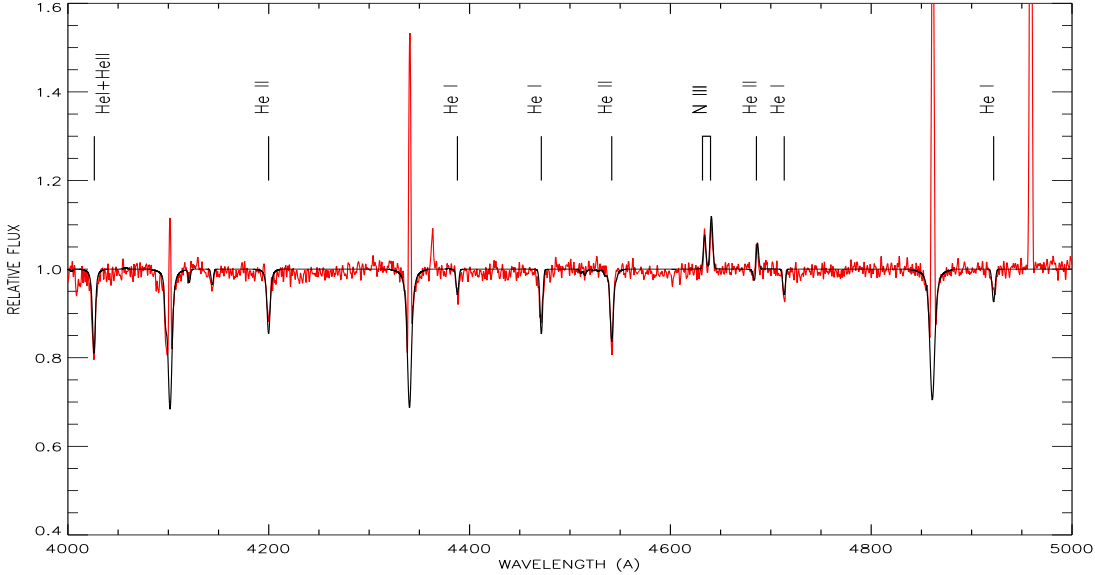}
     \caption{Observed blue spectrum of GHV-62024 (red) and the adopted best fit (black). 
     The main \ion{He}{} and \ion{N}{} lines used in the analysis have been marked. Stellar parameters are given in Table~\ref{modpar}.
     The emission in the red wing of H$_\gamma$ is a sky line.}
	\label{spectrumb}
  \end{figure*}

\begin{figure}
     \includegraphics[angle=90,width=8cm,height=4cm]{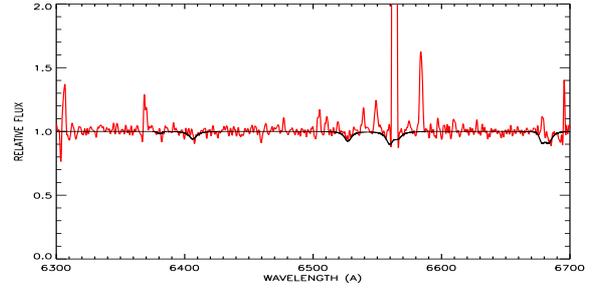}
     \caption{Same as Fig.~\ref{spectrumb}, red spectrum.}
	\label{spectrumr}
  \end{figure}

 Figures~\ref{spectrumb} and~\ref{spectrumr} show the observed coadded blue and red spectra of GHV-62024, respectively (together with the
 best-fit model that we describe below). The stellar spectrum is dominated by the H and He lines, plus a conspicuous 
 \ion{N}{iii} emission at 4630-40 (but without \ion{N}{iv}) and a weak \ion{Si}{iv} 4089 (but \ion{Si}{iv} 4116
 could not be detected). The core of the Balmer lines show an intense nebular contamination, but fortunately the wings are well defined
 (except for H$_\alpha$).
 The red spectrum has a poorer quality, and no individual features could be unambigously identified, apart from H and He lines.
  
The star has a \ion{He}{ii} 4541 line stronger than \ion{He}{i} 4471. By measuring the equivalent widths of these lines
($\log (W^\prime) = \log (EW_{4471})-\log (EW_{4542}) = -0.18$, see~\citealt{conti71}; \citealt{markova11}),  and
comparing with standards we classify the star as O6.5. Because of its \ion{N}{iii} 4634--40 and \ion{He}{ii} 4686 emissions
we assign it the f-class. The luminosity criteria of O stars are based on metal lines that are very faint in our spectrum 
and are usually defined at other metallicities. We assign the star a luminosity class based on the strength of the Balmer lines.
Thus the star is classified as O6.5 IIIf.
The red spectrum is heavily contaminated by nebular and sky emission and consequently H$_\alpha$ yields no 
information on mass-loss rate. Therefore, the
P-Cygni profile at \ion{He}{ii} 4686 is a very important signature. The 19 individual spectra have an average 
SNR of 20 (with a large scatter). 
We then inspected the spectra of individual observing blocks to ensure that the P-Cygni profile is not caused by
unidentified cosmic rays, spikes, or just random noise variations. We also inspected the nebular emission around the
star, and confirmed that there is no emission in \ion{He}{ii} 4686. We conclude that the P-Cygni profile is a stellar feature.
  
\section{Spectral analysis}
\label{anal}

\begin{figure}
     \includegraphics[angle=90,width=6cm,height=4cm]{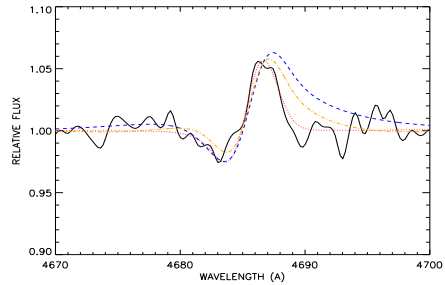}
     \caption{Fit to the \ion{He}{ii} 4686 profile using different values for the terminal velocity and the $\beta$ exponent of the wind velocity law.
     The mass-loss rate was also varied in each case to obtain a reasonable fit while keeping all other parameters fixed.
     Red, dotted line: $\beta= 2.0, V_\infty= 1800~ \mathrm{km~ s^{-1}}, \dot{M}=
      4.3\times10^{-7}~ M_\odot~ \mathrm{yr^{-1}}$ (adopted model); blue, dashed line: $\beta= 1.0, V_\infty= 1800~  \mathrm{km~ s^{-1}}, 
      \dot{M}=  1.1\times10^{-6}~ M_\odot \mathrm{yr^{-1}}$; orange, dot-dashed line:  $\beta= 1.0, V_\infty= 900~ \mathrm{km~ s^{-1}}, 
      \dot{M}=  5.5\times10^{-7}~ M_\odot~ \mathrm{yr^{-1}}$. Black: Observed spectrum.}
	\label{betacomp}
  \end{figure}

We analysed GHV-62024 with the newest FASTWIND version, which includes dielectronic recombination (\citealt{rivero11a}).
The analysis was based mainly on H$_\delta$, H$_\gamma$ and H$_\beta$, \ion{He}{i} 4471, \ion{He}{ii} 4541, 4686 and
the \ion{N}{iii} emission lines at 4634, 4640 \AA. Although all line profiles are sensitive to all parameters, the main diagnostics for gravity
was the wings of the low Balmer lines (except H$_\alpha$ which was not used). For the temperature we mainly used the ratio \ion{He}{i} 4471 to 
\ion{He}{ii} 4541, giving some weight also to the \ion{He}{i} 4387, 4922 lines (which point to a temperature slightly cooler than \ion{He}{i} 4471).
The ratio of \ion{N}{iii} 4634-40 to \ion{N}{iv} 4058 also was a secondary indicator 
(the \ion{N}{iv} line is not present, thus giving an upper limit for the temperature). 
Without a useful H$_\alpha$ profile, the mass-loss rate and the exponent of the velocity law, $\beta$, were determined using 
the \ion{He}{ii} 4686 line. This is crucial, because there is a degeneracy between the $\beta$ exponent and the derived mass-loss rate.
In Fig.~\ref{betacomp} we see the fit to the \ion{He}{ii} 4686 profile using the value of $\beta$= 2.0 and $\beta$= 1.0. From the figure it is clear that the
higher value fits the observed profile better. It turned out to be impossible to find a good fit to the red wing of \ion{He}{ii} 4686 with
a low $\beta$ value. The abundances of \ion{N}{} and He, given in Table~\ref{modpar}, 
were fixed by fitting the profiles of all lines. The first one
is given in the usual logarithmic scale in which the \ion{H}{} abundance is 12, while the \ion{He}{} abundance is given by number, by 
$Y_\mathrm{He}= ( N_\mathrm{He}/N_\mathrm{H})$.
The red spectrum was not used for the analysis, but we note that it is consistent with the adopted parameters (see Fig.~\ref{spectrumr}).

The final stellar parameters are given in Table~\ref{modpar} and the best fits are shown in Figs.~\ref{spectrumb} and \ref{spectrumr}.
For the analysis we adopted a projected rotational velocity of 80 km s$^{-1}$ and a 
radial-tangential macroturbulent velocity of 50 km s$^{-1}$. Although we were unable to determine these values from the line profiles,
at a spectral resolution of $R \approx 2000$ they have little impact on the parameter determination, because the line broadening
is limited by the spectral resolution or the Stark broadening. The adopted metallicity for the model calculations was $Z = 0.15~ Z_\odot$
at the beginning and we changed to $Z= 0.13~ Z_\odot$ for the models close to the final stellar parameters (adopting the
metallicity obtained by \citealt{bresolin07} for B-supergiants, although this difference
in Z produced no appreciable change in the spectrum). 
Unfortunately, there are no features in our spectrum that allow us an independent
metallicity determination. The radius was determined by fitting the visual magnitude by integrating the flux of our final model 
in the Johnson V-filter and after correcting for the distance to IC\,1613 and the reddening (see \citealt{h92} or \citealt{repolust04}). We used
the photometry from \cite{garcia09}, quoted in Tab.~\ref{stardata}. The final model has an intrinsic $(B-V)_0= -0.26$,
while the observed one is $(B-V)= -0.16$. Assuming a standard extinction law with $R_\mathrm {v}= 3.1$ and with the adopted distance modulus 
of 24.31 we finally obtained an absolute magnitude $M_\mathrm{v} = -5.03$, which results in a radius $R= 11.1 R_\odot$ and using
the derived effective temperature, in a luminosity $\log(L/L_\odot)= 5.29$. These values are consistent with a slightly underluminous
O6.5 giant, which reduces the possibility that the star is actually a binary system with two comparable components that could affect the 
spectral analysis.

We have little information on the terminal velocity. We initially adopted $V_\infty/V_\mathrm{esc}= 3.4$ and $V_\infty(Z) \propto Z^{0.12}$ as
given by \cite{castro12} (see also \citealt{castro10}), who used data from the literature (the metallicity dependence has been taken from 
\citealt{leitherer92}). We note that this ratio is higher than the more commonly used one of 2.6,
which would result in a terminal velocity 30$\%$ lower. This has little effect on the analysis, but may be important for the MWM 
(see next section). We adopted a wind terminal velocity of 1800 km s$^{-1}$. This is consistent with the absorption part of the \ion{He}{ii} 4686 P-Cygni profile,
which indicates a lower limit of $\approx$1200 km s$^{-1}$, although a more precise determination of $V_\infty$ from this line is not possible.
The mass-loss rate is derived mainly from the emission profile of \ion{He}{ii} 4686. We obtain ${\dot M}= 4.3\times10^{-7} M_\odot~ \mathrm{yr}^{-1}$. 

We estimated the error in $T_\mathrm{eff}$ to be $\pm1500~ \mathrm{K}$ and in $\log g \pm0.15~ \mathrm{dex}$, which translates into an uncertainty of 
$\pm0.6$ in $R/R_\odot$ and
$\pm0.09$ in $\log(L/L_\odot)$. The spectroscopic stellar mass (the mass obtained using the values of $\log g$ and radius obtained from the
analysis of the spectrum, and corrected from the centrifugal force) results in $M_\mathrm{sp}= 18.3^{+7.8}_{-5.5}~ M_\odot$. The error in the helium abundance is
estimated to be $\pm$0.04 and in $\log (N/H) \pm0.30~ \mathrm{dex}$. The uncertainty in $V_\infty$ is large, $\pm600~ \mathrm{km s^{-1}}$.
The uncertainties in $\beta$ and $\dot{M}$ are more difficult to estimate because the upper limit of $\beta$
(and thus the lower limit of $\dot{M}$) is not well constrained. However, if we consider $\beta$ fixed, the uncertainty in $\dot{M}$ (which is then a
lower limit) is $\pm0.30~ \mathrm{dex}$.

\begin{table*}[htdp]
\caption{Stellar parameters obtained here for GHV-62024 with FASTWIND.  T$_\mathrm{eff}$ is given in K, g in cm s$^{-2}$, $\dot{M}$ in solar masses per year and $V_\infty$ in km s$^{-1}$}
\centering
  \begin{tabular}{lccccccccccc}\hline 
Star ID & Sp Type & $T_\mathrm{eff}$ & $\log g$ & $R/R_\odot$ & $\log(L/L_\odot)$ & $M_\mathrm{sp}/M_\odot$ & $\dot{M}$ & $\beta$ & $V_\infty$ & $Y_\mathrm{He}$ & $\log(N/H)+12$ \\ \hline
62024         &  O6.5 IIIf             & 36500     & 3.60         &  11.1             &  5.29                &  18.3 & 4.30$\times$10$^{-7}$& 2.0 & 1800 & 0.18   & 8.30 \\
\end{tabular}
\label{modpar}
\end{table*}
\begin{table}[htdp]
\caption{Synthetic photometry for the model given in Table~\ref{modpar}. Values are absolute magnitudes.}
\centering
\begin{tabular}{c c c c c c c c}
\hline \hline
U & B & V & R & I \\
\hline
-6.37 & -5.29 & -5.03 & -4.84 & -4.61 \\
\hline
\end{tabular}
\label{modpar2}
\end{table}



%
%
 

\section{The wind momentum of GHV-62024}
\label{discus}
\begin{figure}
\resizebox{\hsize}{!}{\includegraphics{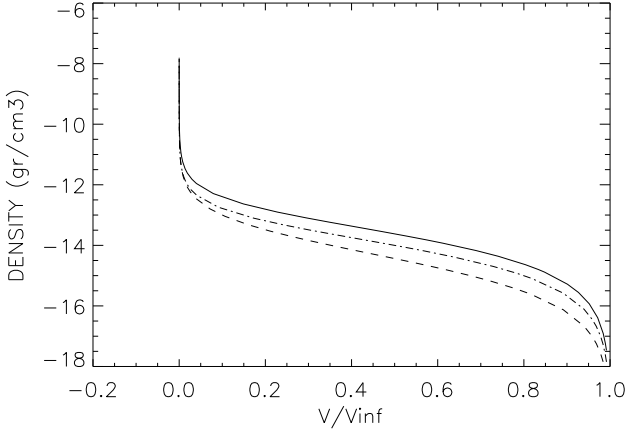}}
\caption{ Run of the density versus wind velocity for three models. The first one (solid line) has 
$\dot{M}= 1.0\times10^{-6}~ M_\odot \mathrm{yr}^{-1}, \beta= 1.0$ and the second one (dashed line)
has $\dot{M}= 4.3\times10^{-7}~ M_\odot \mathrm{yr}^{-1}, \beta$= 2.0. In both cases, the mass-loss rate
is the one need to fit the emission peak of \ion{He}{ii}. The third model (dot-dashed line) has 
$\dot{M}= 4.3\times10^{-7}~ M_\odot \mathrm{yr}^{-1}, \beta= 1.0$. All other parameters are the same as in
our final model.}
\label{vdensity}
\end{figure}

The first remarkable result of the previous section is the high value we obtain for $\beta$, the exponent of the wind velocity law.
The value obtained ($\beta= 2.0$) is much higher than usual values for early and mid Of stars (0.7-1.0, see e.g. 
\citealt{repolust04} for the MW or \citealt{mokiem06} for the SMC). However, $\beta$ and $\dot{M}$ have different
effects on the line profiles. A higher $\beta$ means a slower acceleration in the inner wind (where the \ion{He}{ii} profile
is formed) and therefore, because of the continuity equation
($\dot{M}= 4\pi r^2\rho(r) v(r)$) for a fixed mass-loss rate and terminal wind velocity, at a given point r the velocity is 
lower and the density higher than for a smaller $\beta$ (see Fig.~\ref{vdensity}). This implies more emission at low velocities, 
but also a lower velocity at a given density.  The minimum density required to contribute to the emission profile is reached at
lower velocities in the high $\beta$ model. As a consequence, the line profile in this model has more emission at low velocities compared to
the high velocities. The result is a steeper red wing, fitted with a lower mass-loss rate in the high $\beta$ model.

\begin{figure}
\resizebox{\hsize}{!}{\includegraphics[angle=0]{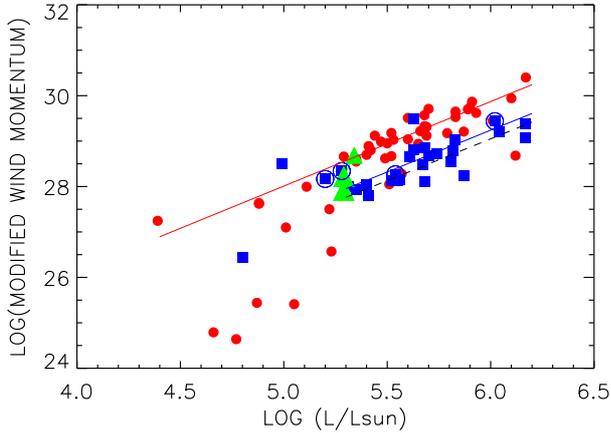}}
\caption{Wind-momentum luminosity relationship for GHV-62024 (upper green triangle: value from \cite{herrero11}; middle green triangle: this work;
lower green triangle: this work, considering a wind terminal velocity 30$\%$ lower than quoted in Tab.~\ref{modpar}) compared to data 
from \citet{mokiem07b} for the MW (red circles) and SMC (blue squares). The SMC stars cited in text were marked with an open circle. From
left to right they are NGC346-010, AzV296, AzV83 and NGC346-001. The solid lines represent the WLR fitted by \cite{mokiem07b} without considering clumping
for MW and SMC objects, and the black dashed line corresponds to the SMC relation shifted to $Z= 0.13 Z_\odot$.}
\label{wlr}
\end{figure}

In a preliminary analysis of GHV-62024, \cite{herrero11}  {\it assumed} a standard value of $\beta$= 0.9 for its wind.
As a result, the MWM rate of this star was much higher than predicted by the \citet{vink01} relationship
for its luminosity and metallicity (note that if the star is contaminated by a companion, which is always a worry at far distances,
the actual luminosity would be lower and thus the discrepancy with the theory larger). 
As indicated in the introduction, this result is similar to that found by \cite{tramper11} for stars in IC\,1613, 
WLM and NGC\,3109, who also adopted $\beta$ (with values from 0.80 to 0.95, depending on luminosity class).
As surprising as the $\beta$ value obtained in this work may look,
it solves the problem of the high MWM of GHV-62024, because it reduces the mass-loss rate while keeping all other parameters
(nearly) fixed. Figure~\ref{wlr} illustrates this point. In the figure we plot the present $\log(D_\mathrm{mom})$ value (28.22) and that obtained by 
\cite{herrero11} (28.69), together with the values given by \cite{mokiem07b} for Galactic and SMC stars. 
The stellar luminosity has also changed because we derive slightly different stellar parameters as a consequence of the new $\beta$.
The new value for the MWM rate fits the cannonical relationship defined by the SMC stars for unclumped winds. We note, however, that because
the metallicity of IC\,1613 is slightly lower than that of the SMC, we could expect the $\log(D_\mathrm{mom})$ of GHV-62024 to be slightly lower as well.

The high $\beta$ value obtained here is unusual, but there are precedents in the literature. The most similar case is probably that of
AzV 83, analysed by \cite{hillier03}. This is an O7 If star (as compared to O6.5 IIIf for GHV-62024) for which the authors also found a velocity law
exponent  $\beta$= 2.0. We do not include clumping in our analysis, but Hillier et al. did, therefore this is not the cause of
the large $\beta$. This high value of $\beta$ for such an early star led the authors to suggest that AzV 83 is a fast rotator seen pole-on.
The peculiar value for $\beta$ would be the result of a mixture of photons from the asymmetric wind\footnote{\citet{hillier03} adopted the
equatorially compressed wind model from \cite{bjorkman93}, i.e., the equatorial wind is denser than the polar one, in contrast to current views,
which favour a a faster and denser wind from the pole and a slower, thinner wind from regions closer to the equator. However, this difference
is not important for the argument presented here.}. Changing acceleration conditions along the wind may mimic a high $\beta$ value (Owocki,
private communication). Extending this argument, we may expect that the equatorial and polar winds are subject to different acceleration
conditions, and accordingly that the global profile mimics a velocity law with a high $\beta$.

There are more similarities between the two stars. \cite{hillier03} obtained $Y_\mathrm{He}= 0.20$ and $12+\log(N/H)= 8.41$, compared to our
$Y_\mathrm{He}= 0.18$ and $12+\log(N/H)= 8.30$. In both cases, the high abundances could be attributed to the proposed fast rotation. The main
difference between both objects lies in the wind terminal velocity. AzV 83 has a modest $V_\infty= 940~ \mathrm{km~ s^{-1}}$ as derived from its
UV spectrum. This value is much lower than the 1800 km s$^{-1}$ we estimated for GHV-62024 by scaling its escape velocity. The difference
would have been smaller had we also used the escape velocity to determine the terminal wind velocity of AzV83. 
The escape velocities of the two stars are not so different: $436~ \mathrm{km~ s^{-1}}$ for AzV83 and $666~ \mathrm{km~ s^{-1}}$
for GHV-62024 (once more, it is true that the devil is in the details). Thus, when estimated from the respective escape velocities, the terminal
velocities of both stars would differ by only $\sim$30$\%$. 

Could it be that the high $\beta$ value derived here for GHV-62024 is the result of a too high estimation for $V_\infty$? 
As was shown by \citet{puls96}, it is primarily the Q-parameter ($ Q= \frac{ \dot{M} }{(V_\infty R_*)^{3/2}}$)
that determines the net wind emission. Therefore, for a given wind emission, if $V_\infty$ increases $\dot{M}$ has to increase as well.
And given the behaviour of density and velocity explained above and illustrated in Fig.~\ref{vdensity}, this could result in too much emission at high velocities.
Could a lower $V_\infty$ result in a lower $\beta$ value? The answer is no (see Fig.~\ref{betacomp}). Although a given stellar parameter has an effect on the
whole profile of a given spectral line, its impact on different parts of the line is also different. Consequently $\beta$ primarily determines the 
steepness of the red wing emission (because a higher $\beta$ means a lower wind acceleration, and therefore the inner layers are denser
and slower, resulting in more emission at a given wavelength but extending to lower velocities). On the other hand,
the main impact of $V_\infty$ is in the extension of the lines profiles, primarily in the absorption part of the P-Cygni profile.
In Fig.~\ref{betacomp} we show an example with different values of $V_\infty$ and $\beta$. We see that we can reduce the red emission by
decreasing the terminal velocity at $\beta$=1.0, but then neither the slope of the red wing nor the extension of the blue absorption would fit.
We conclude that our estimation of the wind terminal velocity cannot be 
wrong by more than 30$\%$, which is consistent with the lower limit indicated in Sect.~\ref{spec}, 
and that this would not affect the derived $\beta$ significantly.

AzV 83 is one of the stars considered by \cite{mokiem07b} (who adopted the 
values given by \citealt{hillier03} after correcting for the clumping adopted by these authors) 
and appears in Fig.~\ref{wlr}, in a position fully compatible with the
rest of the stars. Again, a more standard $\beta$ value would result in a MWM value higher than the average relation.

However, a too high wind terminal velocity would have a clear impact on $\log(D_\mathrm{mom})$. As $\log(D_\mathrm{mom})= \log(\dot{M} V_\infty R^{0.5})$,
using the above definition of $Q$ we obtain $\log(D_\mathrm{mom})= 2.5 \log V_\infty + 2 \log R + \log Q$. For a fixed pair ($Q, R$) we obtain that
a change of 30$\%$ in $V_\infty$ results in a change of 0.3 dex for $\log(D_\mathrm{mom})$. This would bring the point of GHV-62024 in Fig.~\ref{wlr}
close to the value expected for the IC\,1613 relationship. This is what one would expect even for a fast rotator, because \cite{maeder01} has shown that the
WLR (and in turn the position of the star in Fig.~\ref{wlr}) is not altered by fast rotation. 

Therefore, we conclude that these two stars are consistent with a scenario in which they are fast rotators seen pole-on, their wind
momentum is what is expected from their luminosities and metallicities. We suggest as a working hypothesis for future
investigations that the fact that their $\beta$ values are higher than usual may be a consequence of fast rotation.
 
\begin{figure*}
     \includegraphics[angle=90,height=8cm,width=15cm]{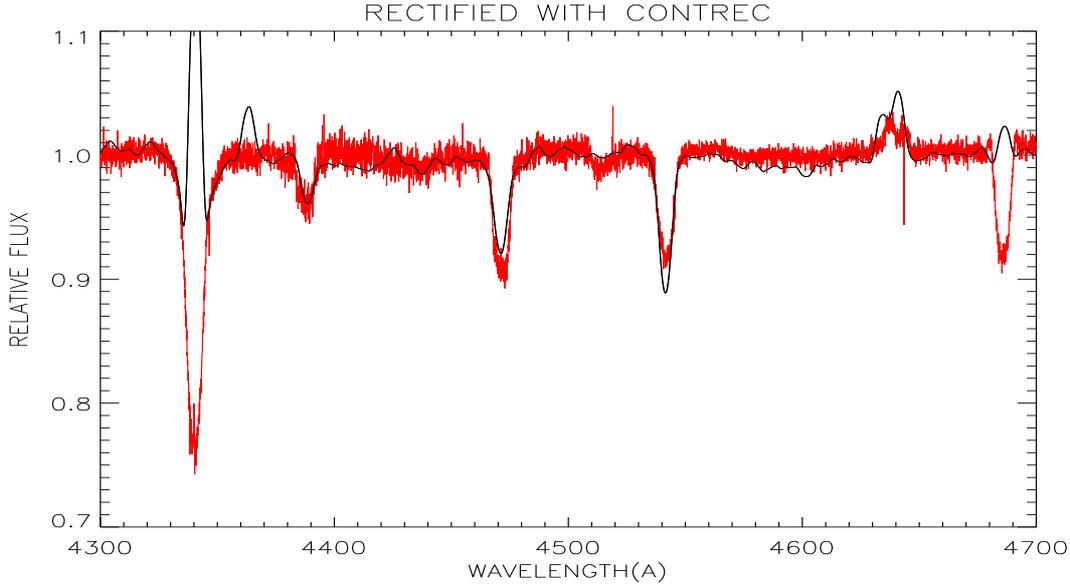}
     \caption{Comparison of the spectra of GHV-62024 (black) and NGC346-010 (red). }
	\label{smccomp}
  \end{figure*}

Can we also observe higher $\beta$ values for known fast rotators?
Two fast rotators in the sample of \cite{mokiem07b} lie at positions close to GHV-62024 in the WLR diagram. 
NGC346-010 is an O7 IIIn((f)) star analysed by \cite{mokiem06}. NGC 346-010 is in fact quite similar to GHV-62024, except for the 
higher mass-loss rate and He abundance of the latter. Fig.~\ref{smccomp} offers a comparison of the spectra of NGC346-010 and GHV-62024, 
with the last one convolved with an additional rotational profile 
with a velocity of 200 km s$^{-1}$ (to simulate the high rotational velocity of NGC346-010, but taking the much lower spectral resolution for GHV-62024 
into account). In the figure we appreciate that both spectra are very similar, except for the emission at \ion{He}{ii} 4686 and  \ion{N}{iii} 4630--40, 
a higher ratio \ion{He}{ii} 4542 / \ion{He}{i} 4471 and the nebular emission in H$_\gamma$. The similarity in the rest of the spectral features indicates
that the GHV-62024 emission in \ion{He}{ii} 4686 is a real feature. This figure can be compared with the spectrum of AzV 83, shown by \cite{hillier03} in their
Fig. 2. There we see the strong emission in \ion{N} 4634--40 and \ion{He} 4686 (this time the line is fully in emission). Thus, the three stars look quite
similar, except for these emission features (and the small differences in the other \ion{He}{} profiles related to effective temperature, which could also be affected
by differences in atmospheric extension).
Because of the lack of UV data, the wind terminal velocity of NGC346-010 
has been again estimated from the escape velocity (with $V_\infty= 1832~ \mathrm{km~ s^{-1}}$ in \citealt{mokiem06}, later revised to 1486 km s$^{-1}$ 
in \citealt{mokiem07b}). A standard value, $\beta= 0.8$, was {\it adopted} for the exponent of the velocity law. For this fast rotator, the
authors suggest the presence of an equatorial disk (because of some difficulties in the line profile fitting, but without other evidence)
although on the whole the fit to the H/He optical spectrum
is satisfactory. The second fast rotator in the sample of \cite{mokiem07b} is AzV 296, a fast rotating O7.5 V((f)) analysed by \cite{massey04}. 
In this case, the terminal velocity has been obtained from UV spectra ($V_\infty= 2000~ \mathrm{km~ s^{-1}}$ ). A $\beta=0.8$ has also been {\it adopted} for the 
exponent of the velocity law. In this case the optical fit is poor and the authors 
suggest that the star might be a binary, which would also explain the relatively large radius needed to fit the observed magnitude and 
the high radial velocity observed.

In both cases, a $\log(D_\mathrm{mom})$ slightly larger than the cannonical relationship is derived. 
Unfortunately, all lines are strong in absorption including H$_\alpha$, and therefore the $\beta$
exponent could not be accurately constrained. Consequently, the mass loss rate in this case may actually be an upper limit. A higher $\beta$ value 
would therefore bring these two fast rotators into better agreement with the average WLR for their metallicity.

Therefore, we conclude that the analysis of these stars indicates that the velocity law may be altered in some stars. Anticipating data from
the next section (where we discuss the \ion{N}{} and \ion{He}{} abundances), we suggest that fast rotation is a probable cause for this
alteration. Note, however, that the evidence is indirect and based on very few objects. Moreover, of
three stars with known "f" signature at low Z (AzV 83 and NGC\,346-001 in the SMC and GHV-62024 in IC\,1613), two 
have been suggested to be seen pole-on. This is quite improbable, unless the f-phenomenon at low Z is favoured by this
pole-on view. If that were the case, the enhanced polar wind would be responsible for the presence of the f characteristics. This possibility agrees
well with the findings by \citet{rivero11a}, who found that the f-phenomenon is related to wind strength and might therefore decrease at low metallicity.

\cite{mokiem07b} have analysed other fast rotators in the SMC, but they give no further clues because the authors could only derive upper limits
that are above the mean WLR for the SMC. However, it is interesting to note that we do not find this situation in the Milky Way: the values derived 
by \cite{mokiem07b} and \cite{repolust04}, whether actual values or upper limits, are close to or below the average WLR for the Milky Way 
and we are
not aware of significantly high $\beta$ values needed to fit the spectra of Galactic O stars, although they are found for the slower winds
of B-supergiants (\citealt{kudritzki99}; \citealt{crowther06}; \citealt{markova08}). We note that determining
$\beta$ from optical spectra of rapidly rotating stars seen edge-on is in most cases very difficult because of the lack of 
spectral features sensitive to this parameter.

We have not taken clumping into account in our analysis and therefore the derived mass-loss rate has to be
considered as an upper limit \citep{puls08}. However, as far as we compared ours with other results and theoretical predictions that
also use unclumped winds, our conclusions remain valid. Of course, there might be relative differences in the clumping among stars with different
properties, in particular the winds of fast rotators might be clumped in a different way than those of slowly rotating stars. However, studies about the
clumping distribution  are at their very beginning, and therefore we have no data yet to discuss this point.


\section{The evolutionary status of GHV-62024}
\label{evol}

\begin{figure}
   \resizebox{\hsize}{!}{\includegraphics[angle=90]{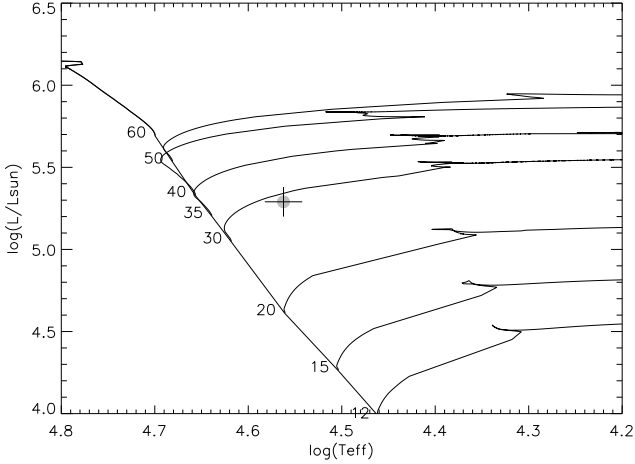}}	
     \caption{Position of GHV-62024 in the HR Diagram. Evolutionary tracks are taken from \citet{brott11}, with an initial rotational velocity of 410 km s$^{-1}$.}
	\label{hrd}
  \end{figure}

In Fig.~\ref{hrd} we place GHV-62024 in the Hertzsprung-Russell diagram and compare it with evolutionary tracks from \citet{brott11} for the SMC.
We chose the tracks with an initial rotational velocity $V_{\rm rot}= 410~ \mathrm{km~ s^{-1}}$ because they have the highest rotational velocity
consistent with the star locus. For higher rotational velocities, the corresponding tracks evolve bluewards already from the ZAMS. 

We see that according to these tracks the star was born with an initial mass of $\sim$30 M$_\odot$. This result does not change significantly when using 
other tracks, even those without initial rotational velocity. The present evolutionary mass derived from the adopted tracks 
(at $V_\mathrm{rot,ini}= 410~ \mathrm{km~ s^{-1}}$)
is $28.6 M_\odot$. A comparison with the spectroscopic mass reveals then the well-known mass-discrepancy.

\begin{figure}
   \resizebox{\hsize}{!}{\includegraphics[angle=90]{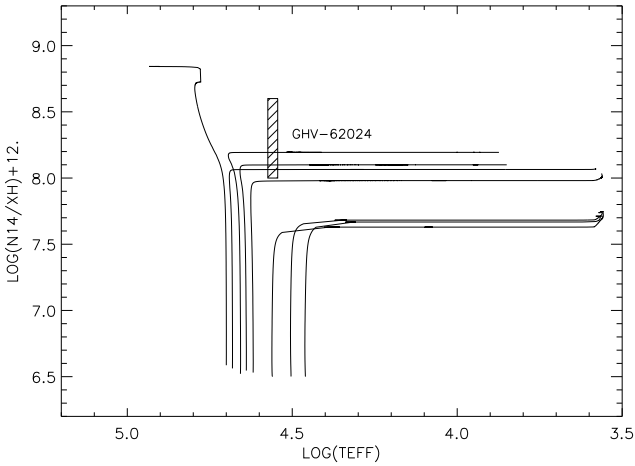}}	
     \caption{Variation of the abundance of $^{14}$N along the stellar lifetime (as given by the evolution in temperature). Represented in the ordinate axis is
     $\log(^{14}N/H)+12$. Models with a zero initial rotational velocity only change their nitrogen abundance in the red supergiant phase, whereas models
     with an initial rotational velocity of $410~ \mathrm{km~ s^{-1}}$ suffer an early nitrogen enhancement. Evolutionary models have masses of 60, 50, 40, 35, 30,
     20, 15 and 12 $M_\odot$, from left to right. The position of GHV-62024 is represented by the dashed box.}
	\label{nabun}
  \end{figure}

The most interesting parameter for the evolutionary status of GHV-62024 is its nitrogen abundance. We obtained a value of 
$\log(N/H)+12= 8.30\pm0.30~ \mathrm{dex}$
from our analysis in Sect.~\ref{anal}. This strong nitrogen enhancement (and the similarly high He abundance, $Y_\mathrm{He}= 0.18\pm0.04$) 
cannot be explained with non-rotating models, which only predict \ion{N}{} and \ion{He}{} enhancement
much later, in the red supergiant phase. Therefore, we looked at the set of rotating models calculated by \citet{brott11} for the SMC metallicity. 
Fig.~\ref{nabun}~ presents the predictions of models with $V_\mathrm{rot}= 410~ {\mathrm {km~ s^{-1}}}$ compared with the value derived from the observations. 
The models predict a slightly lower maximum \ion{N}{} abundance (except for homogeneous evolution) and never reach the derived value of 8.3,
although they are consistent with the lower half of the error bar. We have to take into account that the highest possible N enrichment
is limited by the initial abundances of C and O (which decrease with the metallicity), and that the \ion{C}{} and \ion{O}{} abundances could scale 
differently from solar; alternatively, there might still be some problems in our \ion{N}{iii} line-synthesis \citep{rivero11a}, which might add to the quoted error
budget. The \ion{He}{} abundance is also higher than predicted for a model with an initial mass $M_0= 30~ M_\odot$ at the chosen
initial rotational velocity, but would be consistent with that of a model of $40~ M_\odot$, for which $Y_\mathrm{He}= 0.20$ is predicted at the temperature obatined for
GHV-62024, or with that of a slightly faster rotating model. 

Our results qualititatively agree with those from \citet{rivero11b} for the LMC, who also obtained very high \ion{N}{} and
\ion{He}{} abundances for a large fraction of their LMC sample. 
GHV-62024 would very nicely fit in Rivero-Gonz\'alez et al. Fig. 8 (lower panel), where they show a clear correlation between
the abundances of these two elements. This indicates that in both cases the \ion{N}{} and \ion{He}{} enrichment has the same origin, presumably the
exposition at the surface of the products of the CNO bi-cycle taking place in the stellar interior. 

\begin{figure}
   \resizebox{\hsize}{!}{\includegraphics[angle=90]{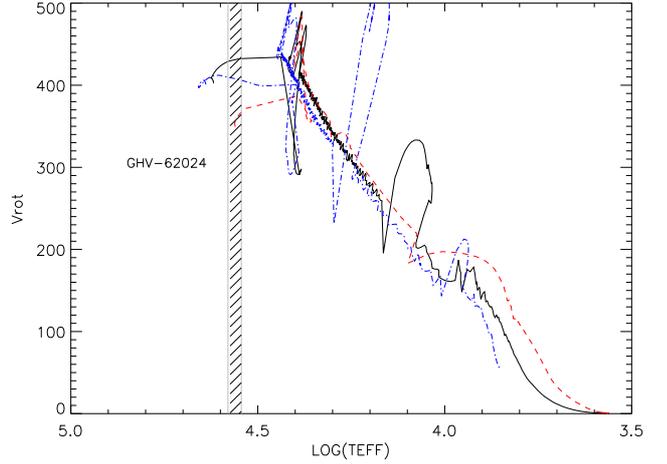}}	
     \caption{Evolution of the rotational velocity for evolutionary models with initial masses of 30 (black), 35 (blue, dot-dashed line) and 20 (red, dashed line) 
     solar masses, according to \citet{brott11}. The temperature range derived for GHV-62024 is indicated by the hatched region.}
	\label{rot}
  \end{figure}

Using the same tracks, we can estimate the stellar age to be less than 4.5 Myr, in qualitative agreement with our expectations for a young region. In Fig.~\ref{rot}
we plot the behaviour of the rotational velocity versus the effective temperature in the models. Interestingly, we see that the rotational velocity is
predicted to remain at high values until much later in the evolution of the star (the effective temperature acts here as a temporal variable). This is fully
consistent with our expectation that the star is a fast rotator. Therefore,
we think that it is possible to conclude that GHV-62024 shows strong \ion{N}{} and \ion{He}{} abundances that are 
compatible with a high initial rotational velocity still present in the star, 
although other possibilities (such as an interactive binary history) cannot be discarded with the present data.

However, if the star still retains its high rotational velocity (note in Fig.~\ref{rot} that for the model with 30 M$_\odot$ the rotational velocity 
increases during the early evolution), this could mean that 2D effects may become important. In that case, our 1D models would be
too simplified and orientation effects could play a role. A proper comparison would have be
made between 2D evolutionary and atmospheric models, which is a work for the future. However, we are confident that even if the high 
rotational velocity is confirmed, the analysis and comparisons described here are a good representation of the average properties of the star.

\section{Conclusions}
\label{conc}
We have analysed GHV-62024, an O6.5 IIIf star in the low-metallicity galaxy IC\,1613. 

Our preliminary analysis of this star \citep{herrero11} resulted in a too high MWM rate for the star, typical of Milky Way metallicity
stars. This would have posed a problem for the theory of radiatively driven winds. However, in our new analysis presented in this work we conclude
that the MWM is consistent with a lower-than-SMC metallicity, but we derive a $\beta$ exponent of 2.0 for the wind velocity law. This is higher than the
usual value for stars of this spectral type. Following a suggestion by \citet{hillier03}, we hypothesize that the star is a fast rotator seen 
from low inclination angle and that the high
$\beta$ would then be an artefact caused by the superposition of light emitted from regions with different wind properties.

When comparing stellar evolutionary models with rotation from \citet{brott11} we see that the star shows the well-known mass-discrepancy.
We derive high N (log(N/H)+12.= 8.30$\pm$0.30) and He (Y= 0.18$\pm$0.04) abundances, which can qualitatively be explained by a strong mixing with processed 
CNO products. The N abundance is higher than the maximum value predicted by the tracks from \citet{brott11} for non-homogeneous evolution, 
although it is consistent within the errors, and the \ion{He}{} abundance is consistent with the model prediction either for a higher mass than the initial
stellar mass derived from the position of the star in the HRD or for a faster initial rotational velocity. However, with an initial rotation
of 410 km s$^{-1}$ the models predict that the star can reach high abundances while still keeping the high rotational velocity. This is therefore
consistent with the above hypothesis that GHV-62024 is a fast rotator seen from a low inclination angle, whose velocity field 
properties would have been modified towards high $\beta$ values without modifying the average WLR
(in agreement with \citealt{maeder01} and \citealt{petrenz00}). 


However, a high rotational velocity could give rise to 2D effects, not included in the 
atmosphere or evolutionary models. If the high rotational velocity is confirmed, the analysis and comparisons carried out in this work should be
considered as a first step towards a more complete, multidimensional analysis.


\begin{acknowledgements}
We want to thank an anonymous referee for very useful comments that helped to improve our paper.
This work was supported by the Spanish Ministerio de Ciencia e Innovaci\'on (grants AYA2008-06166-C03-01 and -02, AYA2010-21697-C05-01 and -04) 
and the Consolider-Ingenio 2010 Program (CSD2006-00070) and the Gobierno de Canarias (grant PID2010119). 
J.G.R.G. gratefully acknowledges financial support from the German DFG, under grant 418 SPA 112/1/08 (agreement betweenthe DFG and the Instituto de Astrofisica de Canarias). 
\end{acknowledgements}


\bibliographystyle{aa}
\bibliography{biblio}

\end{document}